\documentclass[envcountresetchap]{svmult}

\usepackage{graphicx}         
\usepackage[bottom]{footmisc} 

\usepackage{epsfig}

\begin{document}

\setcounter{page}{291}
\setcounter{chapter}{25}

\title
{Offdiagonal complexity:
A computationally quick
network complexity measure.
Application to protein networks and 
cell division\thanks{
Published in:
Mathematical Modeling of Biological Systems, Volume II. A. Deutsch, R. Bravo
de la Parra, R. de Boer, O. Diekmann, P. Jagers, E. Kisdi, M. Kretzschmar,
P. Lansky and H. Metz (eds). Birkh\"auser, Boston, 291-299 (2007).
}}
\titlerunning{Offdiagonal complexity for graphs}

\author{Jens Christian Claussen}
\institute{
Institut f\"ur Theoretische Physik
und Astrophysik, 
Christian-Albrechts-Universit\"{a}t 
zu Kiel, 
Leibnizstr.15, D-24098 Kiel, Germany
\texttt{claussen@theo-physik.uni-kiel.de}
}
\maketitle

\begin{abstract}
Many complex biological, social, and economical networks show
topologies drastically differing from random graphs.
But, what is a complex network, i.e.\
how can one quantify the complexity of a graph?
Here  the Offdiagonal Complexity (OdC),
a new, and computationally cheap,
measure of complexity is defined, based on the node-node link
cross-distribution, whose nondiagonal elements characterize the graph
structure beyond link distribution, cluster 
coefficient and average path
length.
The OdC apporach is applied to the {\sl Helicobacter pylori}
protein interaction network
and randomly rewired surrogates thereof.
In addition, OdC
is used to characterize the spatial complexity of cell aggregates.
We investigate the earliest embryo development states of 
{\sl Caenorhabditis elegans}.
The development states of the premorphogenetic phase
are represented by 
symmetric binary-valued cell connection matrices with
dimension growing from 4 to 385.
These matrices can be interpreted as
adjacency matrix of an undirected graph,
or network.
The OdC approach allows to describe quantitatively the 
complexity of the cell aggregate geometry.
\end{abstract}
\index{complexity}
\index{complexity!of networks}
\index{degree correlations}
\index{offdiagonal complexity}
\index{networks!complex}
\index{networks!random}
\index{C.elegans}
\index{development}
\index{cell division}

\vspace*{-5.5ex}
\section{Complex networks}
From a series of seminal papers  
(Watts \& Strogatz \cite{wattsstrogatz}, 
Barabasi \& Albert
\cite{barabasialbert,albertbarabasi,barabasilinked},
Dorogovtsev \& Mendes  \cite{doromend},
Newman \cite{newman},
see also \cite{bornschu} for an overview)
since 1999,
small-world and
scale-free networks have been a hot topic of investigation
in a broad range of systems and disciplines.
\\
Metabolic and other 
biological networks,
collaboration networks, www, internet, etc.,
have in common
that the distribution of link degrees follows a 
power law, and thus has no inherent scale.
Such networks are termed `scale-free networks'.
Compared to random graphs,
which have a
Poisson link distribution and thus
a characteristic
scale, they share a lot of different properties,
especially 
a high clustering coeff\hspace*{0.001em}icient, and
a short average path length. 
However, the question of {\sl complexity} of a graph
still is  in its infancies. 
A `blind' application of other complexity measures 
(as for binary sequences or computer programs)
does not account for the special properties
shared by graphs and especially scale-free graphs
as they appear in biological and social networks.

Mathematically, a graph (or synonymously in this context,
a network) is defined by a (nonempty) set of nodes, a set of
edges (or links), and a map that assigns two nodes
(the ``end nodes'' of a link) to each link.
In a computer, a graph may be represented either
by a list of links, represented by the pairs of nodes,
or equivalently, by its adjacency matrix $a_{ij}$ whose
entries are 1 (0) if nodes $i,j$ are connected (disconnected).
Useful generalizations are weighted graphs, where the 
restriction of $a_{ij}$ is relaxed from binary values
to (unsually nonnegative) integer or real values 
(e.g.\ resistor values, travel distances, interaction coupling),
and directed graphs, where $a_{ij}$ no longer needs
to be symmetric, and the link from $i$ to $j$ and 
the link from $j$ to $i$ can exist independently
(e.g.\ links between webpages, or scientific citations).
In this chapter 
the discussion will be kept limited to binary
undirected graphs.
\vspace*{-.4ex}

\section{Complexity measures in biology}
\vspace*{-.8ex}
In biological sciences, the evolution of life is studied
in detail and at large; and 
it is observed qualitatively that evolution
creates, on average, organisms of increasing complexity.
If one wants to quantify an increase of complexity,
one has to define siutable complexity measures.
In some sense, the number of cells may be an indicator,
but quantifies rather body size than complexity.
Instead one may observe the number of organelles, the size of 
the metabolic network, the behavioural complexity of social organisms,
or similar properties.
To have a time series of the complexity distribution 
of all organisms during evolution on earth,
would be highly interesting for the
test of models of evolution, speciation and extinctions.
But apart from such academic questions, there are many
areas of practical use of complexity measures
in biology and medicine, 
as the 
complexity of morphological structures, cell aggregates,
metabolic or genetic networks, 
or neural connectivities.
\vspace*{-.4ex}

\section{Other complexity measures}
\vspace*{-.8ex}
For text strings (as computer programs, or DNA)
there are common 
 complexity measures 
in theoretical computer science,
such as {\sl Kolmogorov complexity}
(and the related {\sl  Lempel-Ziv complexity}
and  {\sl algorithmic information content} AIC)
\cite{gellmannLloyd}.
For example, AIC is defined by the length of the shortest
program generating the string.
For random structures, thus also for random graphs, 
these measures indicate high complexity.
A distinction of complex structured (but still partly random)
structures from completely random ones 
usually is prohibitive for this class of measures.
For this reason, measures of {\sl effective complexity}
\cite{gellmann} have been discussed; 
usually these are defined as an entropy (or description length)
of ``a concise description of a set of the entity's regularities''
\cite{gellmann}.
Here we are mainly interested in this second class,
and straightforwardly one would try to apply 
existing measures, e.g., to the link list or to the
adjacency matrix. 
However, mathematically it is not straightforward
 to apply these text string based measures to graphs,
as there is no unique way to map 
a graph onto a text string.

Thus one desires to use complexity measures
that are def\hspace*{0.001em}ined directly 
for graphs.
Two classical
measures are known from graph theory;
 {\sl graph thickness} 
and
 {\em coloring number}
have a low ``resolution'' and 
their relevance
for real networks is not clear.
Two new complexity measures 
recently have been proposed for graphs,
{\sl Medium Articulation}
\cite{wilhelm}
for weighted graphs
(as they appear in
foodwebs)
and 
a measure 
for directed graphs
by Meyer-Ortmanns
\cite{meyer}
based on the 
{\sl network motif} concept 
\cite{alon_motif}).
Unfortunately, the latter two complexity measures are 
computationally quite costly.
A computational complexity approach has been 
defined by Machta and Machta \cite{machtamachta}
as {\em computational depth} 
of an {\em ensemble of graphs}
(e.g.\ small-world, scalefree, lattice). 
It is defined as the number of
processing time steps a large parallel
computer (with an unlimited number of processors)
would need to generate a {\em representative}
member of that graph ensemble.
Unlike other approaches, it does not assign single 
complexity values to each graph,
and again is nontrivial to compute.

Table \ref{tableCOMPLEX} gives a qualitative assessment
of the behaviour of some of the mentioned complexity
measures for lattices in 2D and 3D, complex 
and random structures. Note that especially the ability 
to distinguish nonrandom complex structures from 
pure randomness differs between the approaches.
Hence,  a 
{\sl simpler 
estimator} 
of graph complexity is
desired,
and one possible approach, the
Offdiagonal Complexity,
is proposed here.
A striking observation 
of the node-node link correlation matrices
of complex networks
\cite{claussenddhs03,claussendd}
is, that
entries are more evenly spread among the offdiagonals,
compared to both regular lattices and random graphs.
This can now be used to \mbox{def\hspace*{0.001em}ine} 
a complexity measure, 
 for undirected
graphs
 \cite{claussenddhs03,claussendd}.

This chapter is organized as follows.
In Sec.\ \ref{sec_ODCdef}
OdC is defined and illustrated with an example.
Sections
\ref{sec_helico}
and
\ref{sec_celegans}
investigate the application of OdC to two
quite different biological problems:
a protein interaction network, compared with
randomized surrogates,
and a temporal sequence of spatial cell
adjacency during early {\sl Caenorhabditis elegans} development,
quantifying the temporal increase of complexity.

\begin{table}
\caption{Qualitative assessment of various complexity measures.
\label{tableCOMPLEX}}
\begin{center}
\begin{tabular}{l|c|c|c}
\hline
& 2D, 3D & ~ complex structures ~ & ~ random structures \hspace*{.5mm} \\
\hline
AIC, Kolmogorov 
& $o(1)$ & large & maximal\\
effective complexity ~
& $o(1)$ & large & $o(1)$ \\
coloring number
& 2, 2 \hspace*{1.2mm} & $\simeq 3-4$ & $\simeq 3-4$\\
graph thickness
& ~~ 2, $N^{1/3}$  & $\simeq 2-5$ & $\simeq 3-4$\\
motif count 
& $o(1)$ & large & large\\
Machta 
& $o(1)$ & large & $o(1)$ \\
OdC
& 0 & large & low\\
\hline
\end{tabular}
\end{center}
\end{table}

\clearpage
\section{Definition of the Offdiagonal Complexity (OdC)
\label{sec_ODCdef}}
\vspace*{-0.5ex}
{\bf Definition \rm (Offdiagonal complexity).}
Let $g_{ij}$ be the adjacency matrix of a graph 
with $N$ nodes, i.e., $g_{ij}=1$ if nodes $i$ and $j$ are
connected, else $g_{ij}=0$. 
\begin{description}
\item[(i)]
For each node $i$ of the graph, 
let $l(i)$ be the node degree, 
i.e.\ the number of edges (links),
\begin{eqnarray}
l(i):= \sum_{j=0}^{N-1} g_{ij}
\end{eqnarray}
\item[(ii)]
Let $c_{mn}$
be the number of edges between al pairs
of nodes $i$ and $j$,
with node degrees $m=l(i)$, $n=l(j)$
with $l(j)\geq l(i)$ (ordered pairs), i.e.,
\begin{eqnarray}
c_{mn}:= 
\sum_{j=0}^{N-1}
\sum_{j=0}^{N-1} 
g_{ij}  \delta_{m,l(i)} \delta_{n,l(j)}
H(l(i)-l(j)).
\end{eqnarray}
Here $\delta$ is the Kronecker symbol and 
$H(x)=1$ for $x\leq 0$ and $H(x)=0$ for $x>0$.
Due to the pair odering, the matrix $c_{mn}$ has entries only on
the main diagonal and above.
Thus, $c_{mn}$
is a (not normalized) node-node link correlation matrix.
\item[(iii)]
Summation over the 
minor diagonals, or offdiagonals,
i.\ e.\ 
all pairs with same 
$k_i-k_j$
up to $k_{\rm max}=\min_{i}\{l(i)\}$,
 and normalization, gives us
\begin{eqnarray}
\tilde{a}_{k}= \sum_{i=0}^{k_{\rm max}-k} c_{i,k+i},
~~~~~~~~~~
A:=\sum_{k0}^{k_{\rm max}} \tilde{a}_{k},
~~~~~~~~~~
\forall_k a_k:= \tilde{a}_{k} / A.
\end{eqnarray}
\item[(iv)]
Then OdC is def\hspace*{0.001em}ined as an entropy measure on this 
normalized distributions (here it is understood that $0\ln(0)=0$),
\\[-1.5ex]
\begin{eqnarray}
{\rm OdC } =-\sum_{k=0}^{k_{\rm max}} a_k \ln a_k.
\end{eqnarray}
\end{description}

{\bf Examples:}
For a d-dimensional orthogonal lattice, all nodes have degree $2d$, 
and the node-node link correlation matrix has only one nonzero entry at 
row $2d$ and column $2d$. 
For a 
fully connected graph, the single entry is at row $N$ and column $N$. 
Obviously, 
for regular graphs (where all node have a fixed degree $k$)
OdC=0 holds in general.

\noindent
OdC is an approximative complexity estimator
that takes as values
 zero for a regular 
lattice,
 zero for a fully connected graph,
 low values for a random graph,
and
 higher values for `apparently complex' structures.
One main advantage is that it does not involve 
costly (high-order or NP-complete) computations.

\clearpage
\subsection{Illustration with a spatial network}
\begin{figure}[htbp]
\noindent
\centerline{ \epsfig{file=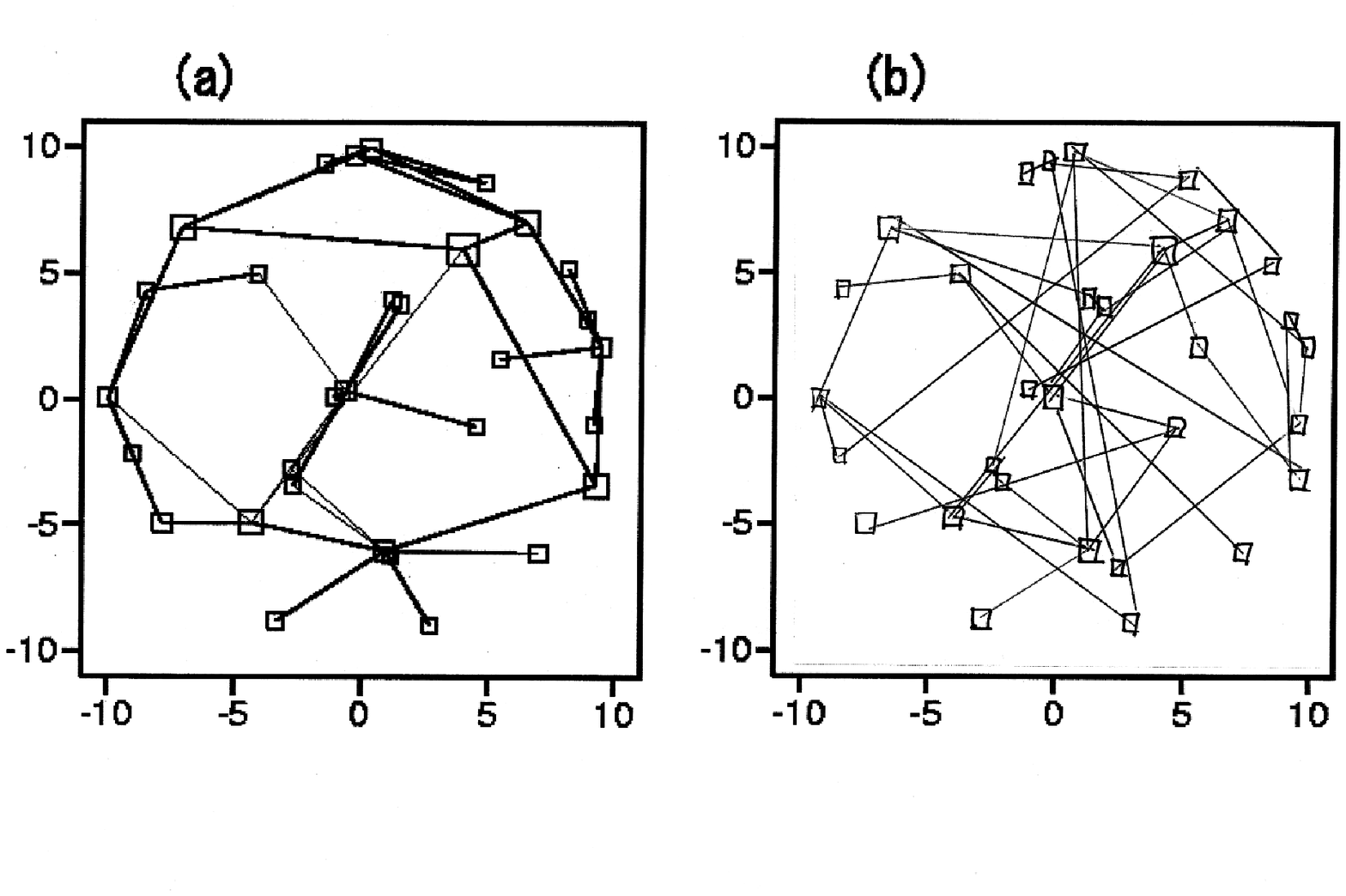,height=4.5cm}
}
\\
{\small
\begin{minipage}[t]{4.3cm}
(a) Self-organized
\\
structure by Sakaguchi
\\
\noindent
\begin{tabular}{cccccccccc}
$k$&1&2&3&4&5&6&7&8\\
$\#k$&10&8&6&4&1&0&1&1
\end{tabular}
\\
Link correlation matrix:
\\
\begin{tabular}{ccccccccccc}
0&0&1&2&0&0&2&5\\
 &3&2&2&2&0&3&1&$\ddots$\\
 & &3&8&0&0&0&1&$\ddots$&5\\
 & & &1&1&0&1&0&$\ddots$&3\\
 & & & &0&0&1&2&$\ddots$&4\\
 & & & & &0&0&0&$\ddots$&0\\
 & & & & & &0&0&$\ddots$&7\\
 & & & & & & &0&$\ddots$&4\\
&&&&&&&&$\ddots$&11\\
&&&&&&&& &7
\end{tabular}
\\
\mbox{The vector of diagonal sums is}
\\
(7,11,4,7,0,4,3,5).
\\
\mbox{{\footnotesize Resulting entropy:} 
${\rm OdC}=1.858622$
}
\\ 
\end{minipage}
\noindent
\hspace*{2cm}
\begin{minipage}[t]{6.3cm} 
(b) Same network, links partly
\\
randomized (1 move/node)
\\
\begin{tabular}{cccccccccc}
$k$&1&2&3&4&5&6&7&8\\
$\#k$&8&7&8&5&2&1&0&0
\end{tabular}
\\
Link correlation matrix:
\\
\begin{tabular}{ccccccccccc}
0&1&4&0&2&1&0&0\\
 &0&7&5&1&0&0&0&$\ddots$\\
 & &2&4&4&1&0&0&$\ddots$&0\\
 & & &3&2&3&0&0&$\ddots$&0\\
 & & & &0&1&0&0&$\ddots$&1\\
 & & & & &0&0&0&$\ddots$&2\\
 & & & & & &0&0&$\ddots$&2\\
 & & & & & & &0&$\ddots$&16\\
&&&&&&&&$\ddots$&15\\
&&&&&&&& &5   
\end{tabular}
\\
The vector of diagonal sums is
\\
(5,15,16,2,2,1,0,0).
\\
{\footnotesize
Resulting entropy:
}
${\rm OdC}=1.376939$  
\end{minipage}
\mbox{}
\\[0.5ex]
\centerline{\small
The random reshuffling lowers the OdC entropy
away from }
\\
\centerline{\small
${\rm OdC}_{max}=2.550838$.
}
}
\caption{
(a) Self- organized structure by Sakaguchi.
(b)
Randomly rewired network.
\label{fig_saka}}
\end{figure}

A spatial hierarchical network emerging from a self-organizing process has 
recently been introduced by Sakaguchi 
\cite{sakaguchi},
as shown in Fig.~\ref{fig_saka}a.
This snapshot example is now taken to illustrate
how the node-node link correlation matrix and
the OdC entropy are modif\hspace*{0.001em}ied under a random
reshuffling of links.
\vspace*{-0.5ex}

\section{Application to the {\sl Helicobacter pylori}
protein interaction graph and reshuffling to a
random graph\label{sec_helico}}
\begin{figure}[htbp]
\centerline{
 \epsfig{file=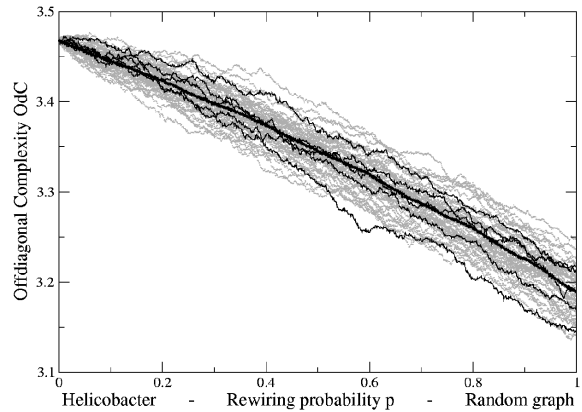,height=5.5cm}
}
\caption{OdC for random reshufflings of the {\sl Helicobacter pylori}
network (left, $p=0$) up to a rewiring probability of $p=1$ (right). 
The bold line shows the average, five OdC trajectories
along a rewiring path are shown for illustration (thin lines).
\label{helico}
}
\end{figure}

\vspace*{-2ex}
To demonstrate that OdC can distinguish between random
graphs and complex networks,
the 
Helicobacter pylori protein interaction graph 
\cite{helico_dat}
has been chosen.
For different rewiring probabilities $p$ and  $10^2$ realizations 
each, the links have been reshuffled, ending up with a random 
graph for $p=1$.
As can be seen in Fig.~\ref{helico},
rewiring in any case lowers the Offdiagonal Complexity.

\section{Application to spatial cell division networks
\label{sec_celegans}}
The tiny  (1mm) nematode worm 
{\sl Caenorhabditis elegans}
looks like a quite primitive organism,
but nevertheless has a nervous system, muscles, 
thus shares functional organs with 
higher-developed animals.
More important, it shows a morphogenetic 
process from a single-cell egg 
thorugh morphogenesis to an adult worm.
Towards an understanding of the genetic mechanisms
of the cell division cycle in general, 
{\sl C.elegans} has become one of the 
genetically best studied animals. 
Despite that, little is known 
(in the sense of a dynamical model) how the
cell divison and spatial reorganization 
takes place. Not even the spatial
organization of cells during morphogenesis 
is well described.

\subsection{Early development of {\sl C.elegans}}
The earliest embryo development states
of
Caenorhabditis elegans
have been recorded experimentally  
and described quantitatively recently
\cite{akraemerphd}.
The cell division development have been described in
simplicial spaces,
and the cell division operations are described 
by operators in finite linear spaces 
\cite{simplicial}.

\subsection{Topological structure during premorphogenesis}
The pre\-mor\-pho\-ge\-ne\-tic phase of development runs until 
the embryo reaches a state of about 385 cells.
The detailed division times and spatial cell movement trajectories
follow with high precision a mechanism prescribed in the genetic program.
While many of the genetic mechanisms are known especially for {\sl C.elegans},
we are a long way towards a mathematical modelling of the cell divison and
spatial organization directly from the genome.
Thus it is still desired to develop 
mathematical models for this spatiotemporal process,
and to compare it with quantitative experimental data.
\\\indent
With good reliability the cell adjacency is known experimentally
\cite{akraemerphd,simplicial}
in a number of intermediate 
steps, which in the remainder we called cell states. 
Here we focus on the adjacency matrices of the cells
describing each intermediate state between cell divisions
and cell migrations,
and investigate the complexity of neighborhood relations.
\\[-6mm]
\subsection{Increasing complexity of C.elegans states}
\begin{figure}[hbtp]
\noindent
\centerline{
\includegraphics[angle=270,width=.464\textwidth]{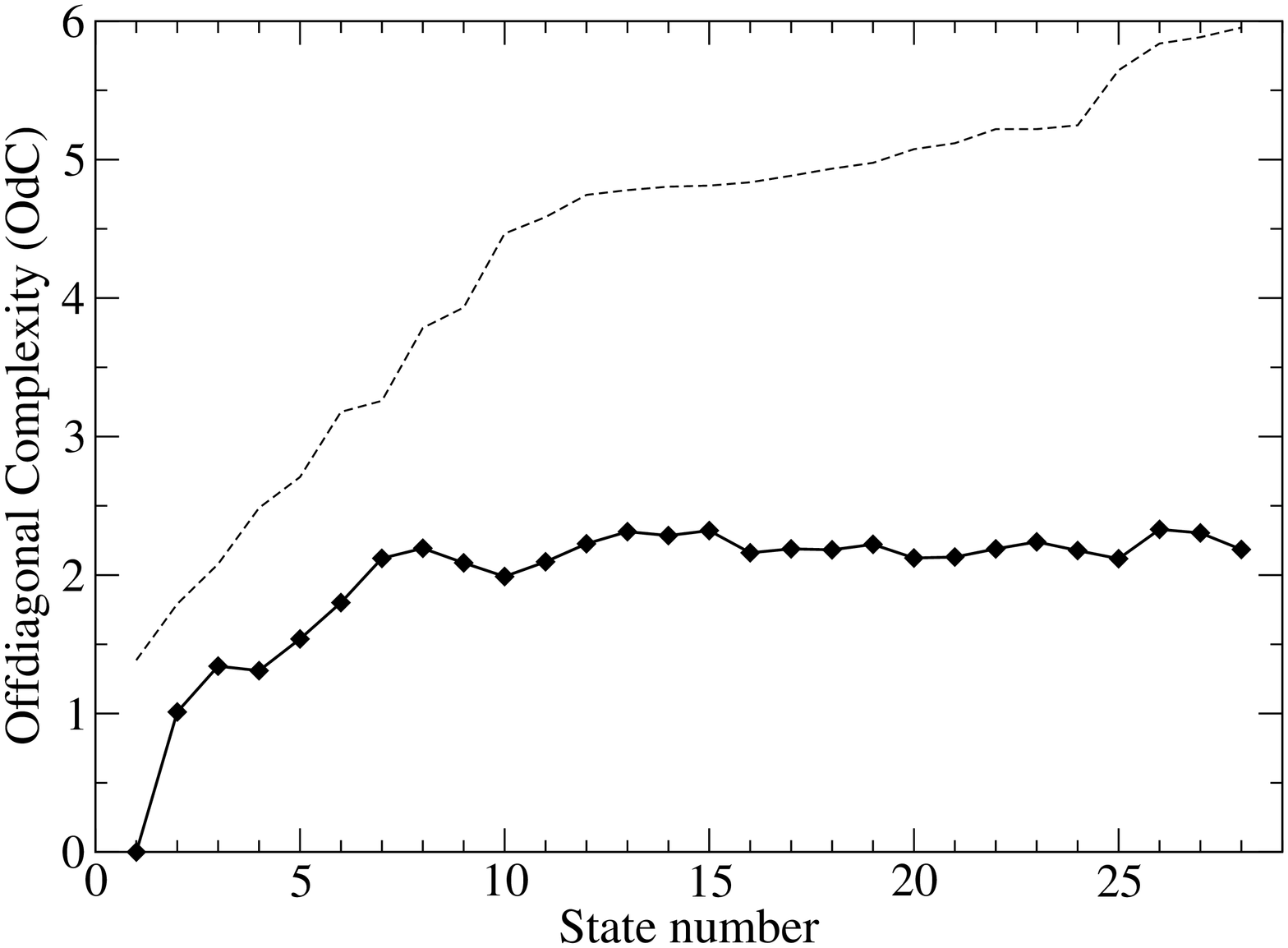}
\hspace*{-0.08\textwidth}
\includegraphics[angle=270,width=.464\textwidth]{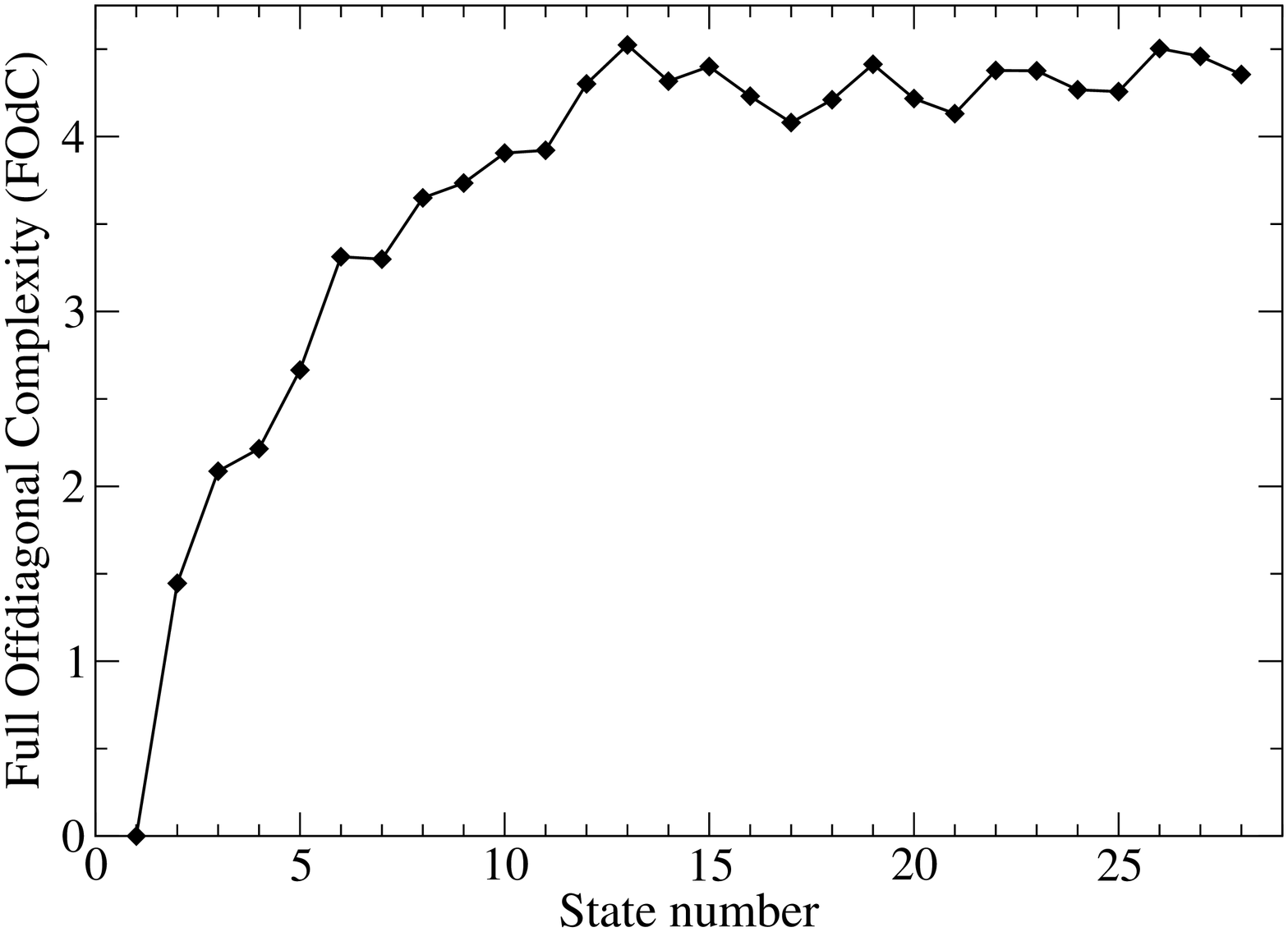}
}\caption{
Left: 
Offdiagonal complexity of the network states. The dashed line
shows the maximal complexity a graph of same number of nodes
could reach. 
Right: 
Full Offdiagonal complexity. Here all possible pairs of nodes
contribute to the complexity.
\label{fig1}
}
\end{figure}

The result for 28 state matrices are shown in
Fig.~\ref{fig1}. The dashed line shows the
supremum value $(-\ln N)$ a graph of the same size could
reach, despite the fact that due to combinatorical 
reasons this supremum is not necessarily always reached.
\\\indent
The moderate decay in the last two states 
may be due to
the fact that 
(at least for Poisson-like link distributions)
the summation implies some self-averaging
if one wants to compare networks of different size.
One way to avoid this problem is to define the 
complexity measure from all $k_{\rm max}\cdot(k_{\rm max}-1)$ entries,
\begin{eqnarray}
{\rm FOdC} := - \sum_{i = 0}^{k_{\rm max}}  \sum_{j=i}^{k_{\rm max}} 
c_{ij} \ln(c_{ij}).
\end{eqnarray}
This can be called Full Offdiagonal Complexity,
as the full set of matrix entries is taken into account.
The result for  FOdC is shown in 
Fig.~\ref{fig1}.
%
\clearpage

\subsection{Saturation for large network size}
\begin{figure}[hbtp]
\noindent
\centerline{
\includegraphics[angle=0,width=.2\textwidth]{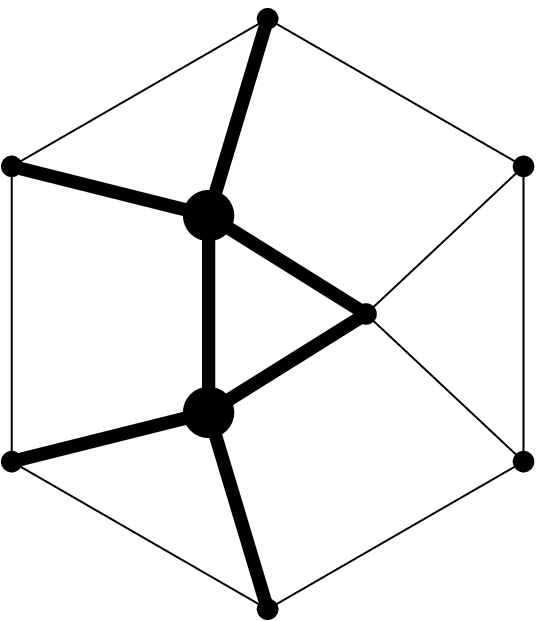}
\hspace*{0.01\textwidth}
\includegraphics[angle=0,width=.2\textwidth]{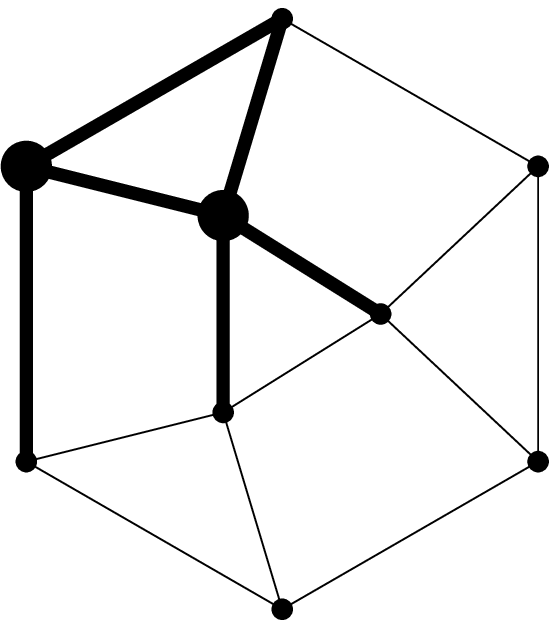}
\hspace*{0.01\textwidth}
\includegraphics[angle=0,width=.2\textwidth]{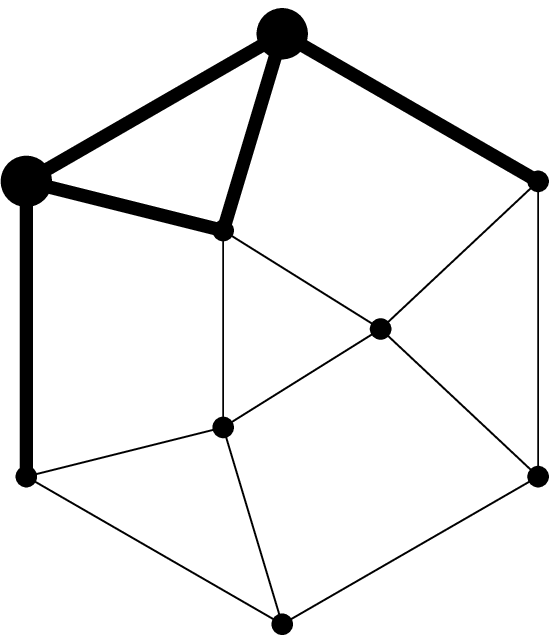}
}\caption{
Intuitive explanation of saturation for large 
homogeneous spatial networks.
From left to right:
Bulk-bulk, bulk-surface, and surface-surface are the
typical pairs of node degrees.
For large cell aggregates, surface and bulk cells are
more homogeneously, i.e.\ the variation of the neighborhood
degree decreases.
\label{fig_spatial2d}
}
\end{figure}

As expected, the complexity of the spatial cell structure
increases along the first premorphogenetic phase.
Compared to the maximal possible complexity that could be 
reached by a graph of same number of node degrees
(but not for a three-dimensional cell complex)
the complexity, as measured by OdC, saturates.
This has a straightforward explanation:
The limiting case of a large homogeneous cell agglomerate would 
end up with roughly  two classes of cells 
(at surface and within bulk) and thus three classes of neighborhood pairs:
bulk-bulk, bulk-surface and surface-surface
(see Fig.~\ref{fig_spatial2d}).
As the coordination numbers within bulk and surface 
fluctuate, this effectively delimits the growth of 
possible different neighborhood geometries.
After initial growth,
FOdC resolves fluctuations corresponding
to the effect of alternating cell division and
spatial reorganization.

\section{Conclusions and Outlook}
\vspace*{-1.5ex}
A new complexity measure for 
graphs and
networks has been proposed.
Contrary to other approaches,
it can be applied to undirected binary graphs.
The motivation of its def\hspace*{0.001em}inition 
is twofold:
One observation is that the
binning of link distributions is
problematic for small networks.
Herefrom the second observation is that 
if one uses instead of the (plain) 
entropy of link distribution,
which is unsignif\hspace*{0.001em}icant for scale-free networks,
a ``biased link entropy'', it has an extremum where the
exponent of the power law is met.
\\
The central idea of OdC is to apply an entropy
measure to  
the 
link correlation
matrix,
after summation over the offdiagonals.
This allows for a quantitative,
yet still approximative, measure 
of complexity.
OdC roughly is  `hierarchy sensitive'
and has the main advantage 
of being
computationally not costly.
\clearpage
\begin{petit}
\section*{Acknowledgments.} 
\vspace*{-2.5ex} J.C.C.\ thanks Christian Starzynski for the simulation code for Fig.~\ref{helico}, and A.\ K{}r\"amer for kindly providing the experimental data  of the cell adjacency matrices.
\end{petit}
\mbox{}\\[-10ex]

\bibliographystyle{plain}

\end{document}